\documentstyle[sprocl]{article}

\def\Journal#1#2#3#4{{#1} {\bf #2}, #3 (#4)}

\def\NPB{{\em Nucl. Phys.} B}
\def\PLB{{\em Phys. Lett.}  B}
\def\PRL{\em Phys. Rev. Lett.}
\def\PRD{{\em Phys. Rev.} D}

\def\be{\begin{equation}}
\def\ee{\end{equation}}
\def\bea{\begin{eqnarray}}
\def\eea{\end{eqnarray}}

\begin{document}


\title{FLAT DIRECTIONS IN SUSY GUTS
\footnote{Talk given by Borut Bajc at SUSY'01, Dubna, Russia}}

\author{BORUT BAJC}

\address{J. Stefan Institute, 1001 Ljubljana, Slovenia. 
E-mail: borut.bajc@ijs.si}

\author{ILIA GOGOLADZE}

\address{International Center for Theoretical Physics,
Trieste, Italy, and Andronikashvili Institute of Physics, 
Georgian Academy of Sciences, 380077 Tbilisi, Georgia. 
E-mail: iliag@ictp.trieste.it}

\author{RAMON GUEVARA}

\address{Dept. of Physics, University of Trieste, 34100 
Trieste, Italy. \\ E-mail: guevara@ts.infn.it}

\author{GORAN SENJANOVI\' C}

\address{International Center for Theoretical Physics,
Trieste, Italy. \\ E-mail: goran@ictp.trieste.it}


\maketitle

\abstracts{It is shown that a realistic SUSY SU(6) GUT can 
dynamically generate the GUT scale and solve at the same time 
the doublet-triplet splitting problem. The cosmological 
implications of such a model are briefly reviewed.}

\section{Introduction}

One of the main reasons to consider supersymmetry as a 
phenomenological viable model is the hierarchy problem: 
why is the Higgs boson so light, since it is naively 
believed that it should get a mass of the order of the 
next new scale, for example the grand unified scale or
the Planck scale. Supersymmetry solves technically this 
problem, assuring that the boson-fermion cancellations in 
loop diagrams do not change considerably the scalar mass. 
There are however two remnants of the hierarchy problem 
still in susy models. 

The first one is to understand, why mass scales in nature 
are so different, for example, why is 
$M_{GUT}\ll M_{Planck}$, or why is $M_{W}\ll M_{GUT}$. 

The second one is how to achieve the mass splitting between 
Higgs SU(2) doublets and Higgs SU(3) triplets, which 
inevitably appear in any GUT. 

This short paper deals with the above problems, as well as 
with cosmological implications. For a longer version see~\cite{our}. 

\section{Flat directions}

It has been shown long ago \cite{witten} that flat directions 
in supersymmetric models can generate dynamically very different 
scales. The idea of such a mechanism, known also as dimensional 
transmutation or radiative symmetry breaking, is very simple: 
a) have a supersymmetric model 
with at least one flat direction; this flat direction is exact 
at any order of perturbation thery due to a supersymmetric 
non-renormalization theorem; b) break supersymmetry spontaneosusly 
or softly, in order to lift this flat direction; higher loop 
contributions to the effective potential can generate a nontrivial 
minimum; since these corrections are logarithmic in the 
flat direction field, this new minimum is exponentially far 
away from the original scale (for example the susy breaking scale 
or the soft mass term). In such a way one can reproduce with 
soft supersymmetry breaking the electroweak~\cite{agpw} or the 
GUT~\cite{tuy} scale from the Planck scale. 

The simplest example of a GUT flat diection is given by

\begin{equation}
\label{w1}
W=\lambda\;Tr(\Sigma^3)\;,
\end{equation}

\noindent
where $\Sigma$ is an adjoint of the gauge SU(6). The flat 
direction is 

\begin{equation}
\label{sol1}
\Sigma=\sigma diag(1,1,1,-1,-1,-1)\;.
\end{equation}

\noindent 
Such simple flat directions are present only in SU(2n) theories.

The solution (\ref{sol1}) with arbitrary nonzero $\sigma$ breaks 
the original SU(6) gauge group to SU(3)$\times$SU(3)$\times$U(1). 
To break further one can use for example the Fayet-Iliopoulos 
D-term of an extra (anomalous) gauge U(1):

\begin{equation}
\label{dterm}
D_{U(1)}=q_H|H|^2+q_{\bar H}|\bar H|^2-\zeta\;,
\end{equation}

\noindent
where $H$ and $\bar H$ are the fundamental and antifundamental 
representations of SU(6) and $q_H$, $q_{\bar H}$ have the same 
sign as $\zeta$. The supersymmetric solution 

\begin{equation}
\label{sol2}
H=\bar H=\sqrt{\zeta/(q_H+q_{\bar H})}\;(0,0,0,0,0,1)
\end{equation}

\noindent
together with (\ref{sol1}) clearly breaks the GUT 
into the standard model. 

\section{The doublet-triplet splitting}

The previous example has unfortunately light Higgs triplets and 
heavy doublets, which is the opposite of what we want. To 
obtain the right pattern we need~\cite{our} at least two adjoints 
($A$, $A'$) and two singlets ($S$, $S'$):

\begin{equation}
W=\lambda\;Tr(A^2A')+\lambda_SS\;Tr(AA')+\lambda_{S'}S'\;Tr(A'^2)\;.
\end{equation}

\noindent
The solution is given by $A$ in the SU(4)$\times$SU(2)$\times$U(1) 
direction and proportional to an undetermined (flat direction) $S$
and a zero $A'$. Using again the D-term (\ref{dterm}), one gets 
exactly two light Higgs doublets and heavy triplets~\cite{dp}.

\section{Cosmology}

At high enough temperature the $-T^4$ term dominates 
and the symmetry restoration takes place \footnote{
An opposite behaviour is however possible due to 
different initial conditions~\cite{dkbs}.}, since a large nonzero vev 
of the flat direction would diminish the number of light degrees of 
freedom~\cite{pietal}. Due to the flatness of the potential, the 
critical temperature~\cite{yamamoto} is only of order 
$(m_{3/2}M_{GUT})^{1/2}$, which is safe for the production of 
monopoles~\cite{our}. Also, the phase transition is first 
order, reducing further the production rate. 
Of course, all this is relevant only if the phase 
transition really takes place, which is the subject of 
further studies.

\section*{Acknowledgments}

We thank the organizers of SUSY'01 for great hospitality and 
extremely well organized conference. We are grateful to Lotfi 
Boubekeur for collaboration in the early stage of this work and 
to Gia Dvali for important comments. The work of B.B. is supported 
by the Ministry of Education, Science and Sport of the Republic 
of Slovenia; the work of I.G and G.S. is partially supported by 
EEC under the TMR contracts ERBFMRX-CT960090 and HPRN-CT-2000-00152. 
Both B.B. and R.G. thank ICTP for hospitality during the course 
of this work.

\end{document}